\documentclass[runningheads]{llncs}
\usepackage[T1]{fontenc}
\usepackage{graphicx}
\usepackage{amsmath}
\usepackage{booktabs}
\usepackage{multirow}
\begin{document}
\title{Orientation recognition and correction of Cardiac MRI with deep neural network}

\author{Jiyao Liu\inst{1} 
}
\authorrunning{J Liu \& K Zhang et al.}
\titlerunning{Orientation recognition and correction of Cardiac MRI}
\institute{Institute of Science and Technology for Brain-Inspired Intelligence, Fudan University, Shanghai 
\\  \email{liujy22@m.fudan.edu.cn
}}

\maketitle              %

\begin{abstract}

In this paper, the problem of orientation correction in cardiac MRI images is investigated and a framework for orientation recognition via deep neural networks is proposed. For  multi-modality MRI, we introduce a transfer learning strategy to transfer our proposed model from single modality to multi-modality. We embed the proposed network into the orientation correction command-line tool, which can implement orientation correction on 2D DICOM and 3D NIFTI images. Our source code, network models and tools are available at \url{https://github.com/Jy-stdio/MSCMR\_orient/}

\end{abstract}
\section{Introduction}

When cardiac magnetic resonance (CMR) images are recorded in DICOM format and stored in a PACS system, they can be stored in different image orientations. Recognizing and understanding this difference is critical for downstream tasks based on deep neural networks (DNNs), such as image segmentation \cite{seg1} and myocardial pathology analysis \cite{anas1}, as current DNN systems typically treat only the inputs and outputs of matrices, without considering imaging orientation and real-world coordinates. This work aims to provide a study of CMR image orientation 
about human anatomy and real-world standardized coordinate systems and to develop an effective orientation recognition and correction method.

Deep learning-based methods have been widely used in orientation recognition and prediction tasks. Wolterink et al proposed an algorithm that extracts coronary artery centerlines in cardiac CT angiography (CCTA) images using a convolutional neural network (CNN) \cite{8}. Duan et al combine a multi-task deep learning approach with atlas propagation to develop a shape-refined bi-ventricular segmentation pipeline for short-axis CMR volumetric images \cite{3}. Based on CMR orientation recognition, we further develop a framework for standardization and adjustment of the orientation.

This work is aimed at designing a DNN-based approach to achieve orientation recognition and correction for multiple CMR modalities. Fig. ~\ref{net} presents the pipeline of our proposed method. The main contributions of this work are summarized as follows:

(1) We propose a scheme to standardize the CMR image orientation and categorize all the orientations for classification.

(2) We present a DNN-based orientation recognition method for CMR images and transfer it to other modalities.

(3) We develop a CMR image orientation adjust tool embedded with a simplified orientation recognition network, which facilitates the processing of large amounts of MRI data.

\section{Method}

In this section, we introduce the proposed cardiac MRI orientation recognition and correction method. We propose a network framework and embed it into the CMR Orientation Correction command-line tool.

\subsubsection{orientation category}

\begin{figure}
\includegraphics[width=\textwidth]{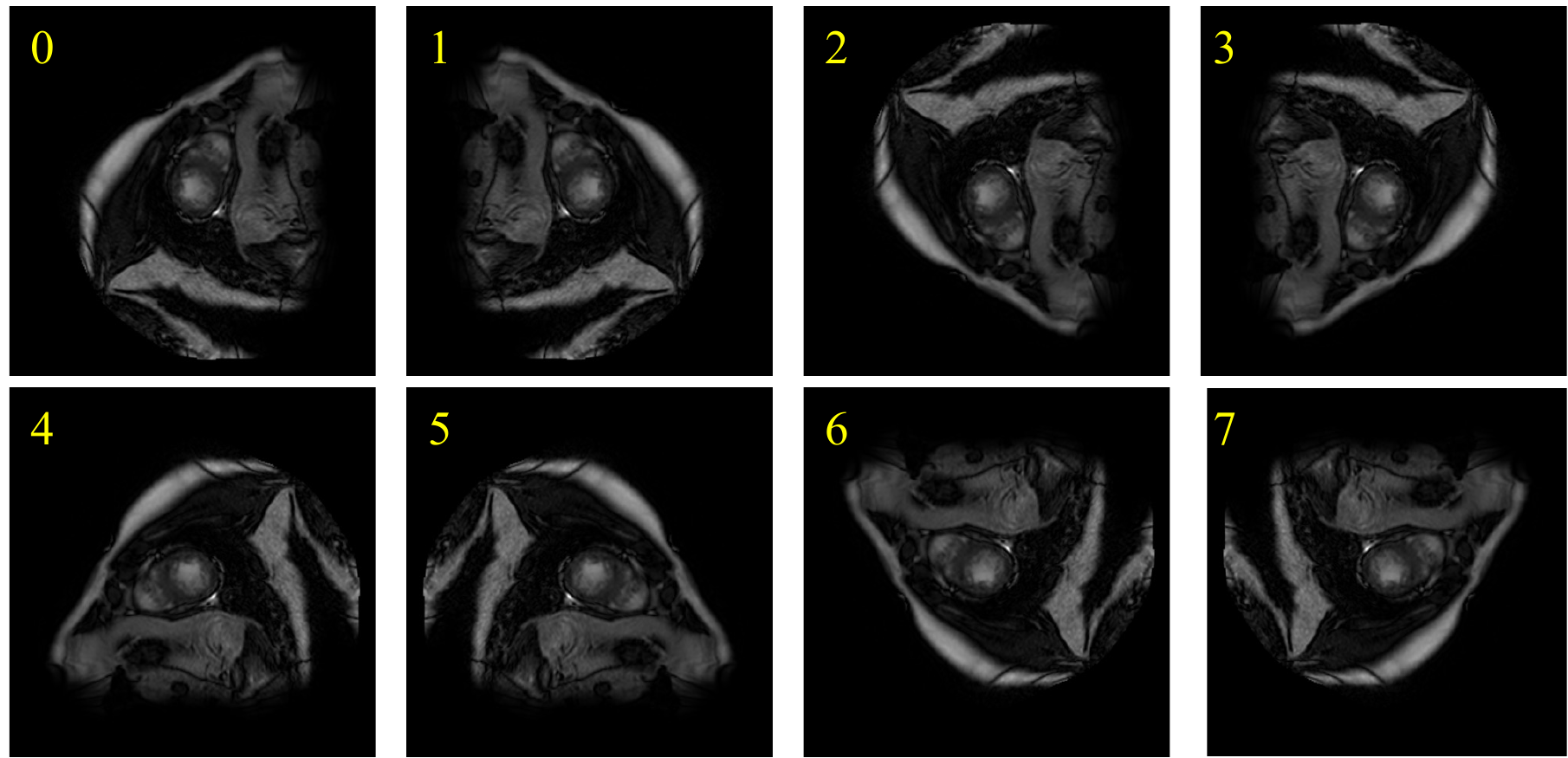}
\caption{The 2D slice of CMR images of the 8 Orientation class.} \label{class}
\end{figure}

Due to differences in equipment and scanning habits, there may be many situations in the Orientation corresponding to the cardiac MRI image, which may cause problems for downstream tasks such as segmentation or detection. Taking a 2D image as an example, we define the four corners of the 2D image in the standard direction as $\begin{bmatrix} 1 & 2 \\ 3 & 4 \end{bmatrix}$, and then the direction of the two-dimensional MRI image may have the following eight changes, which is listed in Table \ref{tab:addlabel}. For each image-label pair $(X_t, Y_t)$. Select a target direction $O_k$ from 8 direction classes, and flip the direction of $X_t$ to the direction of $O_k$. Then get the image-label pair $(X_t^{'}, Y_t^{'})$.The 2D slice of CMR images of the 8 orientations we defined are shown in Fig.~\ref{class}.

\begin{table}[htbp]
  \centering
  \caption{Orientation category of 2D CMR Images, in which sx, sy, and sz respectively represent the axial size of the initial state image, and x, y, and z represent the coordinates of any pixel point on the x-axis, y-axis, and z-axis of images of the other Orientation.}
  \setlength{\tabcolsep}{2mm}%
    \begin{tabular}{cp{9.335em}cc}
    \toprule
    class & \multicolumn{1}{c}{Operation} & Image & Correspondence of coordinates \\
    \midrule
    0     & \multicolumn{1}{c}{initate state} & $\begin{bmatrix} 1 & 2 \\ 3 & 4 \end{bmatrix}$  & Target[x,y,z]=Source[x,y,z] \\
    1     & \multicolumn{1}{c}{horizontal flip} & $\begin{bmatrix} 2 & 1 \\ 4 & 3 \end{bmatrix}$  & Target[x,y,z]=Source[sx-x,y,z] \\
    2     & \multicolumn{1}{c}{vertical flip} & $\begin{bmatrix} 3 & 4 \\ 1 & 2 \end{bmatrix}$  & Target[x,y,z]=Source[x,sy-y,z] \\
    3     & \multicolumn{1}{c}{Rotate 180} & $\begin{bmatrix} 4 & 3 \\ 1 & 2 \end{bmatrix}$  & Target[x,y,z]=Source[sx-x,sy-y,z] \\
    4     & \multicolumn{1}{c}{Flip along the main diagonal}  & $\begin{bmatrix} 1 & 3 \\ 2 & 4 \end{bmatrix}$  & Target[x,y,z]=Source[y,x,z] \\
    5     & \multicolumn{1}{c}{Rotate 90° clockwise} & $\begin{bmatrix} 3 & 1 \\ 4 & 2 \end{bmatrix}$  & Target[x,y,z]=Source[sx-y,x,z] \\
    6     & \multicolumn{1}{c}{Rotate 270° clockwise} & $\begin{bmatrix} 2 & 4 \\ 1 & 3 \end{bmatrix}$  & Target[x,y,z]=Source[y,sy-x,z] \\
    7     & \multicolumn{1}{c}{Flip along the subdiagonal} & $\begin{bmatrix} 4 & 2 \\ 3 & 1 \end{bmatrix}$  & Target[x,y,z]=Source[sx-y,sy-x,z] \\
    \bottomrule
    \end{tabular}%
  \label{tab:addlabel}%
\end{table}%

\subsubsection{recognition network}
\begin{figure}
\includegraphics[width=\textwidth]{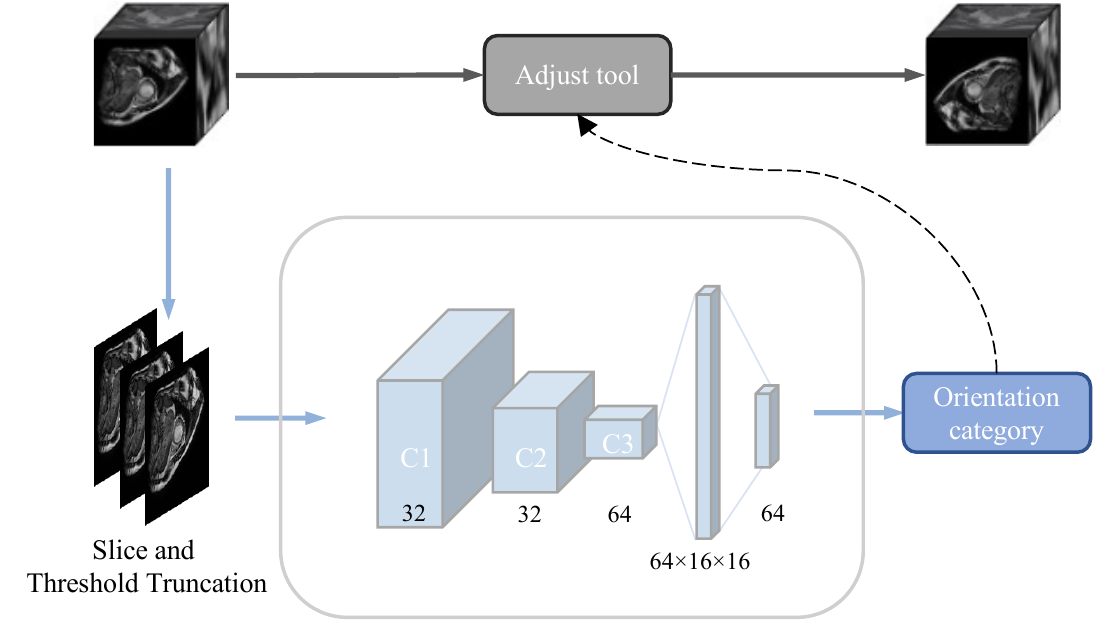}
\caption{The pipeline of the proposed CMR orientation recognition and correction method. The image is first truncated at three gray value thresholds, and after applying histogram equalization, the three-stage images are concated into a three-channel input. We embed the orientation recognition network into the orientation adjustment command-line tool.} \label{net}
\end{figure}

The network architecture we propose consists of two basic components as Fig.~\ref{net}. A Backbone network that consists of 3 layers of CNN to generate feature maps. 2) a classification head of 2 fully connected layers that perform classification based on the aggregated feature maps provided by the two previous components. For the backbone we initially evaluated many different architectures including ResNet\cite{resnet} and DenseNet\cite{dense}. Since the choice of the backbone architecture did not have a significant impact on the overall performance of the final model, we focus on our 3 layers of CNN in this work, because of its efficiency.

Training is conducted in two stages:

- \textbf{preprocess and data augmentation}: The 3D CMR image is sliced into 2D slices, and each slice is resized to (256,256). We define the maximum pixel value of each 2D slice as $X_{max}$, and then three truncation operations are performed on $X_t$ at the threshold of 60\%, 80\% and 100\% of $X_{max}$ to obtain $X_{1t},X_{2t},X_{3t}$. Setting different thresholds enforces the characteristics of the image under different gray value window widths to avoid the influence of extreme gray values. We apply histogram equalization for $X_{1t},X_{2t},X_{3t}$, which is found that it can make the model converge more stably during training.We denote the concatenated 3-channel image $[X_{1t}, X_{2t}, X_{3t}]$ as $X'$. During the training phase, we perform random small-angle rotations, random crops, and resize for the input images. We also apply z-score normalization for the $X'$ to speed up model training convergence.

- \textbf{network training}: The network was trained with 40 epochs, batch size 32, SGD optimizer with default parameters, batch normalization, ReLU activation functions, and a learning rate of $10^{-2}$ for balanced-Steady State Free Precession (bSSFP) cine dataset, respectively.

When adapting the proposed orientation recognition network from a single modality to other modalities, we employ a transfer learning approach to obtain transfer models. For example, we pre-train the model on the bSSFP cine dataset, and then transfer the model to the late gadolinium enhancement (LGE) CMR or T2-weighted CMR dataset. On the new modality dataset, we first load the pre-trained model parameters. We freeze the network parameters of the backbone and retrain the fully connected layers on the new modality dataset. We then proceed to the next fine-tuning step, retraining both the backbone and fully-connected layers until the model converges. In fine-tuning training, we obtain an adapted model on the new modality dataset. The backbone and the whole network were trained with a learning rate of $10^{-3}$ and $10^{-4}$.

The command-line tool is suitable for the visualization and orientation correction of CMR images. It supports batch orientation correction operations of CMR images and provides a simple parameter setting method. By specifying a folder, one line command is enough to identify the orientation of all MRI files in the folder and correct the wrongly oriented files.

\section{Experiment}

We evaluate our proposed orientation recognition network on the MyoPS dataset \cite{PAMI,Spr2016}, which provides the three modalities CMR (LGE, T2, and bSSFP) from the 45 patients. For the simplified orientation recognition network, we train the model for a single modality on the MyoPS dataset, then transfer the model to other modalities. We divide all slices into two subsets, i.e., training set and validation set, with proportions of 80\% and 20\%, respectively.

On an NVIDIA Tesla V100 16GB GPU, the training time is about 15 minutes. Table \ref{tab:result} shows the average accuracy on the test set. Compared with ResNet18 and DenseNet121, our proposed model has sufficient accuracy, and fewer parameters, enabling fast inference. Besides,high-precision, small-parameter supported model results make the tools we develop more efficient.

\begin{table}[htbp]
  \centering
  \caption{Test performance of the comparison methods and proposed approach: bSSFP is pretrained, LGE and T2 is trained via transfer learning.}
  \setlength{\tabcolsep}{4mm}
    \begin{tabular}{lcccc}
    \toprule
          & \multicolumn{3}{c}{\textbf{Accuracy}} & \multicolumn{1}{c}{\multirow{2}[4]{*}{\textbf{\#Parameter}}} \\
\cmidrule{2-4}          & bSSFP & LGE   & T2    &  \\
    \midrule
    ResNet18    & 0.9974    & 0.9986    & 0.9936    & 11.69M \\
    DenseNet121    & 0.9957    & 0.9937    & 0.9923    & 7.98M \\
    Our proposed model & 0.9982 & 0.9914 & 0.9861 & 1.08M\\

    \bottomrule
    \end{tabular}%
  \label{tab:result}%
\end{table}%

\section{Conclusion}

We have proposed a multi-modality CMR orientation recognition and correction network. Additionally, we have developed the CMR Orientation Adjustment command-line tool (CMRcorrect), which is embedded in an orientation recognition network. Experiments demonstrate that the orientation recognition network has an outstanding performance for orientation recognition and correction on multi-modality CMR images. Our future research goals are to refine CMR image orientation classification and cross-modal orientation recognition without fine-tuning.

\end{document}